# Modeling Variability in Template-based Code Generators for Product Line Engineering

Timo Greifenberg[1], Klaus Müller[1], Alexander Roth[1], Bernhard Rumpe[1], Christoph Schulze[1], Andreas Wortmann[1]

**Abstract:** Generating software from abstract models is a prime activity in model-driven engineering. Adaptable and extendable code generators are important to address changing technologies as well as user needs. However, they are less established, as variability is often designed as configuration options of monolithic systems. Thus, code generation is often tied to a fixed set of features, hardly reusable in different contexts, and without means for configuration of variants. In this paper, we present an approach for developing product lines of template-based code generators. This approach applies concepts from feature-oriented programming to make variability explicit and manageable. Moreover, it relies on explicit variability regions (VR) in a code generator's templates, refinements of VRs, and the aggregation of templates and refinements into reusable layers. A concrete product is defined by selecting one or multiple layers. If necessary, additional layers required due to VR refinements are automatically selected.

## 1 Introduction

Engineering complex software systems introduces a conceptual gap between the problem domains and the solution domains of discourse [FR07]. Model-driven engineering (MDE) aims to bridge this gap by lifting abstract models to primary development artifacts. Deriving executable software from models requires extensive handcrafting or code generators. Thus, generating software from abstract models is a prime activity in MDE and many domains have adopted code generation [RR15].

Although reuse is of essence in software engineering, most code generators are monoliths developed for a very specific purpose (such as a certain target platform with specific features) that do not consider reuse or variability as their primary focus. Reusing such code generators in different contexts with different requirements or features is hardly feasible and thus impedes code generator development. One approach to handle variability in such monolithic code generators is to create code generator variants via informal reuse [Jö13] such as copy-paste. In this scenario, the original code generator variant is copied and all required changes are applied to the copy of the variant. The main downside of this approach is that generator changes might need to be applied to all generator copies. This is laborious and error-prone. An alternative to that is to use specific code generation frameworks with built-in support for handling variability [Ac15, Xt15]. Even though this alternative does not result in monolithic code generators, the resulting code generator variants are bound to a specific code generation framework, which might not be feasible for legacy code generators. Additionally, the provided approaches rely on language specific approaches for implementing variability, e.g. design patterns.

---

[1] RWTH Aachen University, Software Engineering, Germany, http://www.se-rwth.de

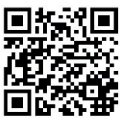

The goal of this paper is to present an approach to develop code generator product lines (CGPLs), which is explicitly designed to handle variability in code generators and which can be applied to any code generator framework. To implement variability in code generators, the approach is based on explicit variability regions (VRs) and the aggregation of templates into reusable layers. Each VR can refine one or multiple VRs from a different layer. A concrete code generator variant is configured by selecting one or multiple layers. In addition to that, further layers are selected automatically, if this is required by the VR refinements. The resulting layers are composed to create a concrete variant. For defining and selecting layers, we provide two domain-specific languages (DSLs).

This idea is rooted in feature-oriented programming (FOP), an implementation technique from classical software product line (SPL) development [Ap13]. We extend the notion of FOP layers [SB98] over templates and define how (parts of) templates can be reused with these layers. The benefit of applying ideas of FOP to CGPL development is that the underlying concept is decoupled from specific template languages and can be applied to any code generator.

In the remainder, Section 2 introduces the variability concepts our approach relies on and Section 3 describes the product configuration mechanisms for code generators. Afterwards, Section 4 demonstrates the application of our approach to a code generation framework. Then, we compare our approach to the informal (copy-paste) approach for creating CGPLs in a case study in Section 5. Subsequently, related work is presented in Section 6 and, finally, Section 7 concludes this contribution.

## 2   Variability Concepts in Code Generator Product Lines

Code generator product lines and common SPLs are both founded on a set of components that are used to create a concrete code generator product or a software product [CN12, PBL05, RR15]. The main difference is that a code generator product is a SPL on its own, since it generates a variety of software products that are similar, and thus shares generator components potentially in different variants [BS99]. As in SPLs, a concrete code generator product, which is referred to as a variant, is a set of selected components with additional adaptations and customizations.

Feature-oriented programming (FOP) [Ap13] is an approach to implement SPLs that is based on building software systems by composing features. A feature represents a configurable unit of a software system that represents a requirement-satisfying design decision [ALS06, Ap13]. Each feature is arranged in a layer [SB02, BSR03, ALS05] that contains artifacts. In order to reuse existing functionality and to successively add new features by adapting existing artifacts, an artifact may refine multiple other artifacts [Ap13]. In FOP, a refinement adds new code to an existing artifact, e.g., adds a new variable to a Java class. Figure 1 shows an example for a stack of three layers with refinements.

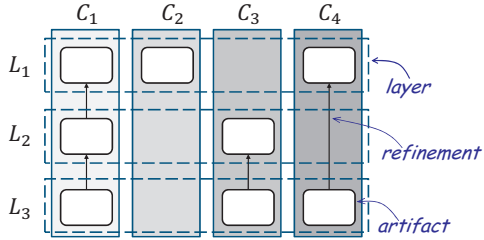

Fig. 1: Example for layers: three vertical layers $L_1$ to $L_3$ refine four artifacts $C_1$ to $C_4$.

In this example, the first layer contains three artifacts. The second layer contains a refinement for the artifact $C_1$ and a new artifact $C_3$. Finally, the third layer contains a refinement of the refinement for $C_1$, a refinement for $C_3$ and a refinement of $C_4$ from the first layer. By merging layers, different variants of a software system are formed. As the used layers contain code generator artifacts, the layers are subsequently called *code generator layers (CGLs)*. As shown in this example, FOP relies on artifact refinements. This is feasible for object-oriented languages but becomes challenging for template languages, as they may differ inherently. Thus, a concept for applying FOP to template languages is needed.

### 2.1 Variability Regions and Variability Region Refinements

*Variability regions* (VRs) provide a template language independent approach to apply concepts of FOP to code generators. A VR represents an explicitly designated region in an artifact that has to be uniquely addressable by an appropriate signature. We distinguish between three types of VRs. First, variability regions are explicitly marked in some way and contain content within an artifact. This, for instance, allows to group a designated part of a template as one VR, which can be refined. Second, variability regions are explicitly marked but are empty, i.e. do not contain any content. Such VRs can be used for future extensions. Third, the complete artifact is regarded as one VR.

For each VR, we define three different refinement operations. First, a `replace` operation completely replaces a VR with some other content. In this case, a certain VR is provided that substitutes the original VR. For example, template code for a Java method can be replaced with a new implementation. Second, content can be added before a VR and, third, content can be added after a VR. Semantically, `before` and `after` mean that specific content should be included before or after a VR. This shares many phenomena with aspect-oriented programming (AOP) [Ki97] applied on templates.

When dealing with `replace` operations, the effect of a `replace` operation to the content added before and after a VR has to be addressed. In this work, VRs are simply replaced but the before and after content, which may have been added, is kept. When the content that replaces a VR or that is added contains VRs, the new content with the VRs is regarded as a complete unit with all replacements and `before` and `after` operations. In consequence, all existing before and after contents have to be composed.

## 3    Code Generator Variant Configuration and Generation

A CGPL consists of a number of CGLs and each CGL contains a number of templates. Before a concrete product of a CGPL can be generated, it has to be defined which refinement operations are performed in which CGL, i.e., which VR contained in a template from a CGL refines which VR contained in a template from another CGL. Based on such a definition of the refinement operations, a concrete code generator variant of the CGPL can be configured and finally generated.

### 3.1    Layer Definition

In our approach, all files encapsulated in a concrete CGL of the CGPL are stored in a specific file system directory, comparable to the work in [ALS05]. The refinement operations that are performed for each CGL are modeled in one layer definition model. To define a layer definition model, we provide a simple domain-specific language (DSL) called Layer Definition Language (LDL). Using LDL, it can be defined which CGL refines which other CGL and which concrete refinement operations are performed. LDL allows for modeling the three refinement operations we introduced in Section 2.1:

- `A replaces B`: The VR with signature `B` is replaced by the VR with signature `A`.
- `A before B`: The VR with signature `A` is added before the VR with signature `B`.
- `A after B`: The VR with signature `A` is added after the VR with signature `B`.

An example for a layer definition model defined using LDL is shown in List. 1. At first, this example states that CGL `factoryVariant` refines CGL `baseVariant` (l. 1). Subsequently, the layer definition model defines which concrete refinement operations are performed (ll. 2−3). As explained in Section 2.1, we require each VR to be uniquely identifiable by its signature. In List. 1, the first refinement operation (l. 2), which is a `replace` operation, refers to the signatures `EntityExt:AdditionalMethods` and `ClassMain:Methods`. By default, each VR signature starts with the path to the artifact containing the VR (relative to the CGL directory) and its name. Hence, the first refinement operation expresses that the artifact `EntityExt` contains a VR `AdditionalMethods` and that this VR replaces the VR with name `Methods` contained in artifact `ClassMain`. Signatures for VRs can also be constructed in different ways, as long as it is possible to uniquely identify the artifact and the VR in that artifact. The second refinement operation (l. 3) states that the VR `ClassCopyright`, which represents a complete template, is added before the VR `Class`, which represents a complete template too. If the CGPL contains other CGLs with refinement operations, these have to be defined in the layer definition model too.

As this example already indicates, layer definition models are not restricted to particular types of languages. The only decision that has to be made is how to uniquely identify a VR within an artifact written in a particular language.

```
                                                              LayerDefinition
1 layer factoryVariant refines baseVariant {
2   EntityExt:AdditionalMethods replaces ClassMain:Methods;
3   ClassCopyright before Class;
4 }
```

List. 1: A layer `factoryVariant` that defines two refinements of the layer `baseVariant`.

### 3.2 Variant Configuration

Based on the layer definition model, a concrete generator variant can be configured by defining which CGLs of the CGPL should be selected. As a consequence of this, a concrete generator variant will be created which results from composing the VRs of the selected CGLs with the VRs they refine and all other not refined VRs of the selected layers. This procedure is repeated for the refined CGLs until a CGL is traversed which does not refine any other CGL. To configure a concrete generator variant, we define a product configuration model using a simple DSL called Product Configuration Language (PCL). In PCL, the name of the resulting concrete generator variant has to be defined and it has to be stated which CGLs should be selected. In each PCL, at least one CGL must be selected. Moreover, it can optionally be defined into which output folder the artifacts of the resulting generator variant are written.

An example for a product configuration model defined in PCL is shown in List. 2. According to this configuration, the resulting generator variant will be called `FactoryGenerator` and this variant is constructed by selecting the CGL `factoryVariant`. Moreover, the artifacts of `FactoryGenerator` would be written to the output folder gen. To infer which CGLs need to be composed to create the `FactoryGenerator`, the layer definition model needs to be analyzed. In this example, the layer definition model is given in List. 1 and it indicates that the CGL `factoryVariant` refines CGL `baseVariant`. Thus, both CGLs need to be composed to create `FactoryGenerator`. However, before a concrete generator variant can be composed, it has to be ensured that the layer definition model is valid, i.e., a set of layers can be computed and VR refinements are unambiguous. Validation is required to ensure that the selected code generator product can actually be build.

```
                                                                  ProductCfg
1 generator FactoryGenerator {
2   output = "gen";
3   layers = "factoryVariant";
4 }
```

List. 2: Example for a product configuration model selecting layer `factoryVariant`

To validate the layer definition model, we map it to colored directed graphs, where each vertex represents a VR, each edge a refinement, and the color represents the layer a VR belongs to. First, the refinement operations for the selected layers are processed. For each refinement two vertices are introduced, if they are not already existing in the graph: one for the refining VR and one for the refined. The added vertices represent the VRs with all their contained VRs. Additionally, a directed edge between the two vertices is created. It points from the refining VR to the refined. Each vertex that represents a VR of the current layer is colored in a particular color, that represents the layer. The other vertex is colored in another color that represents the other layer. After processing the refinement operations of the selected layers, the graph is traversed. Each time a new vertex with a color that has not yet been processed is found, the layer definition model is processed as described above.

A layer definition model does not induce any conflicts if and only if: 1) for any two vertices $v_i^q$ and $v_j^c$ with colors $q$ and $c$ and and a path $(v_i^q, v_j^c)$ there exists no other path $(v_j^c, v_i^q)$, i.e., the graph does not contain a cycle and 2) there exists no other vertex $v_f^g$ with color $g$ such that $(v_f^g, v_j^c)$ holds, i.e., a VR is not refined by multiple VRs. A cycle in the graph makes it impossible to perform the composition automatically, as it results in an infinite loop of refinements with no dedicated end point. Furthermore, if multiple CGLs are selected and in this selection, a VR is refined by multiple VRs, it cannot be automatically decided which of these multiple refinements should actually take place in the composition. In both situations, it is necessary to resolve the problem manually. To derive a valid configuration, we can employ any graph traversal algorithm and select the different colors of the visited vertices, which represent the different layers.

In Figure 2, an example for a graphical representation of a layer definition model is given on the left-hand side. The resulting graph structure is shown on the right-hand side. It is assumed that artifacts of layer $L_1$ are colored in purple ($p$), artifacts of $L_2$ are colored in orange ($o$) and artifacts of $L_3$ are colored in blue ($b$). We further assume, that the layer $L_3$ is manually selected and, therefore, layers $L_2$ and $L_1$ are automatically selected because of their refinements. In this example, there is a path $(C_1^b, C_1^p)$ as $L_3$ contains a refinement of a refinement of $C_1$ from $L_1$. In addition, as $L_2$ is automatically selected, the refinements $(C_2^b, C_2^o)$ and $(C_2^o, C_2^p)$ produce a conflict.

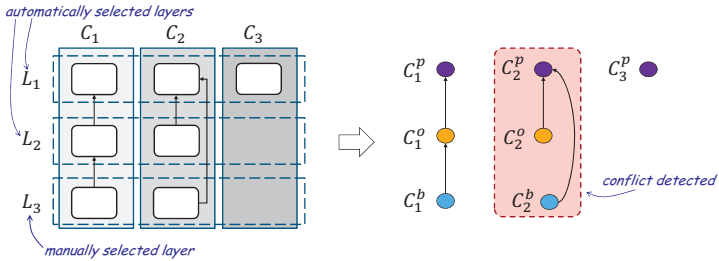

Fig. 2: A three-layered example, where layer $L_3$ is manually selected and layers $L_2$ and $L_1$ are automatically selected because of their refinements. This selection introduces a cycle and, thus, a conflict.

### 3.3 Building Variants by Composition

Based on the layer definition model and the product configuration model, a concrete code generator variant is created. To this effect, the templates, and corresponding refinements are composed. A template that is not refined in one of the relevant layers is used as is for the resulting code generator variant.

Refining templates requires proper application of refinement operations prior to variant composition. To compose a variant, we first need to (a) transitively determine all layers that have to be additionally selected because of the refinements and (b) compose the resulting layers to define the code generator variant.

In general, there are two options on composing VRs. The first option is to perform composition at run-time of the generator, called *generation-time*. In this case the VR operations are executed at generation-time. This means that no VRs are created but support for generation-time execution of VR operations is required. As an alternative, the VRs can be composed by creating new VRs which contain the composition results. The main question that needs to be answered when applying this latter approach is how to deal with `before` and `after` operations. To avoid that the generator framework has to be extended to be able to handle these operations at generation-time, on template level, `before` and `after` operations can be replaced by template inclusion statements in the according template language.

The composition of two layers means that all contained artifacts and their VR refinements are composed. Two layers are composed if and only if there is at least one VR refining a VR in the opposite layer, according to the understanding of composition as defined for FOP [Ap13]. If more than two layers are involved in the composition, then we process all refinements for one VR sequentially in a bottom-up way. If in this sequence a refinement is a `replace` operation, then the VR being replaced is substituted by the VR replacing it. Moreover, if a refinement in a sequence denotes a `before` or `after` operation, then the refining VR is added before (respectively after) the refined VR.

An algorithm for performing this composition would start visiting all selected layers and then the automatically added layers. In each layer, every refinement is considered in a bottom-up way, i.e., only the outgoing refinements refining a VR are considered.

## 4  Demonstrating Example for Variability Regions

In this section, we demonstrate the application of our approach to the code generator framework openArchitureWare using Xpand [Xp15] as a template language. Motivated by an industrial use case (see Section 5), the openArchitectureWare framework in version 3.0.1 has been chosen.

### 4.1  Example Description

In this example, we consider a code generator that processes a class diagram (CD) as input and translates every class into a Java class with the same name. Each attribute of a class is translated to a Java variable with a mutator and an accessor method. It also adds a public constructor with an argument list containing all attributes defined in the class. For demonstration purposes, we assume the input CD contains the class `Person` with the attribute `name` of type `String`. On the left-hand side, Figure 3 shows a CD of the resulting generated class.

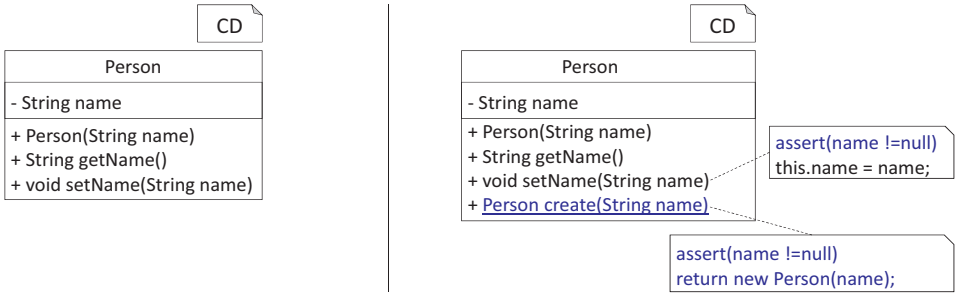

Fig. 3: Overview of the originally generated output (left) and the required output (right).

Another context requires to generate the code for classes differently. Instead of writing a new code generator from scratch or copying the original one, a new variant of the existing generator should be created. This variant should validate the argument passed to the mutator methods and produce a factory method that asserts proper creation arguments. A CD of the resulting output is depicted in Figure 3 on the right-hand side. Here, the generated class name corresponds to the input model's class name, it features an assertion in `setName()`, and provides a factory method `create()` for `Person` objects that asserts that the value passed for name actually exists. Please note that, usually the constructor visibility would be changed too, to prevent others from invoking the public constructor directly. However, due to limitations of space, we omitted this part and assumed that the constructor visibility is not changed. To achieve this kind of extensibility, our approach lifts the code generator to a CGPL by explicitly managing variability.

### 4.2   openArchitectureWare

The Xpand template language allows to split templates into multiple blocks. Such blocks begin with the keyword `DEFINE` and, thus, we henceforth refer to these as `DEFINE` blocks. Each `DEFINE` block is identified by a name and is defined for a specific type of input model element, called *meta model class*. For instance, all concrete classes of the CD in our example are represented by the meta model class `MMClass`.

List. 3 shows an excerpt of the realization of our example in Xpand. For the sake of brevity, only those parts are shown that are relevant for the refinement of the template. The first `DEFINE` block with name `ClassImpl` is defined for the meta model class `MMClass`. If this `DEFINE` block is invoked for a concrete class, a new Java file is created for that class, indicated by the `FILE` statement (l. 2). Expressions encapsulated in [...] lead to the invocation of the according methods of the meta model classes. The results of these invocations are inserted into the output at the current location.

To generate the implementation for a class, the `DEFINE` blocks `Constructor` (l. 4) and `FurtherMethods` (l. 6) are invoked for the class and `SetterMethod` (l. 5) is invoked for each attribute of the class by using the `EXPAND` statement. The string that is constructed in the according `DEFINE` blocks is inserted into the output at the current location.

```
 1 [DEFINE ClassImpl FOR MMClass]
 2   [FILE ... Name".java"]
 3   class [Name] {
 4     [EXPAND Constructor]
 5     [EXPAND SetterMethod FOREACH Attribute]
 6     [EXPAND FurtherMethods]
 7   }
 8   [ENDFILE]
 9 [ENDDEFINE]
10
11 [DEFINE Constructor FOR MMClass]
12   public [Name](...) {
13     [EXPAND ConstructorImpl]
14   }
15 [ENDDEFINE]
16
17 [DEFINE SetterMethod FOR MMAttribute]
18   public void set[UpperCaseName]([Type] [Name]) {
19     [REM]BEGIN VR:SetterMethodBody[ENDREM]
20       this.[Name] = [Name];
21     [REM]END VR:SetterMethodBody[ENDREM]
22   }
23 [ENDDEFINE]
24
25 [DEFINE FurtherMethods FOR MMClass]
26 [ENDDEFINE]
```

List. 3: Template `Class` (in Folder `base`) showing an excerpt of the base template for the translation of CDs into Java code realized with Xpand.

### 4.3 Mapping Variability Regions to Templates

In Section 2, we introduced three kinds of VRs: non-empty VRs that refer to a particular region within an artifact, empty VRs for future extensions and the VR representing the complete artifact. In Xpand, non-empty VRs can be introduced by defining non-empty DEFINE blocks and, accordingly, an empty VR can be introduced by declaring an empty DEFINE block. The most important aspect of every VR is that it has to be uniquely identifiable through its signature. The signature of a DEFINE block can be derived by the path to the template and its name. If multiple DEFINE blocks with the same name exist in one template, the meta model class of such a DEFINE block has to be stated in the signature as well. Otherwise, it cannot be differentiated between the different blocks with the same name. In addition, the complete template represents a VR as well with the path to the template and its name representing the signature of this VR.

Besides interpreting every DEFINE block as one VR, it is possible to introduce further VRs into Xpand templates explicitly by using, e.g., named comments around the corresponding region in the template. The advantage of using comments for this is that the template language does not have to be extended and this approach is applicable to all template languages supporting comments. This approach is comparable to utilizing protected regions for integrating handwritten and generated code [Gr15], as there, comments mark the regions into which handwritten code can be inserted. List. 3 shows an example in lines 19 to 21, in which the body of the setter method is contained in the VR `SetterMethodBody`. The start and the end of the VR `SetterMethodBody` is denoted through comments, represented by `REM` and `ENDREM`, in which the name of the VR is defined. This comment-based approach is used here only for demonstration purposes, to illustrate how it can be applied. When using Xpand, instead, a separate `DEFINE` block could have been used as well. Even though this approach allows for introducing any kind of VR into a template, this approach is rather fragile, as a comment can be changed by accident easily. Moreover, a template might contain several other comments which makes it more difficult to identify VRs marked by comments.

### 4.4 Variability Region Refinements

Using the template introduced in List. 3, we show how VR refinement operations can be mapped to concepts in Xpand. This is done by using the layer definition model shown in List. 4. Moreover, List. 5 illustrates the refining template used for the example.

```
1 layer factoryVariant refines baseVariant {
2    base.ClassWithFact:FurtherMethods
3       replaces base.Class:FurtherMethods;
4
5    base.ClassWithFact:Method.SetterMethodBody
6       replaces base.Class:SetterMethod.SetterMethodBody;
7 }
```
*LayerDefinition*

List. 4: Layer definition model for Xpand realization.

As indicated by the first refinement operation, the VR `FurtherMethods`, contained in template `ClassWithFact` (ll. 1-8 of List. 5) which is located in folder `base`, replaces the empty VR `FurtherMethods` from template `Class` (ll. 25-26 of List. 3), which is located in folder `base` too. By means of this, the factory method `create()` is generated additionally.

Furthermore, the VR `SetterMethodBody` contained in the `DEFINE` block `Method` in template `ClassWithFact` (ll. 11-14 of List. 5) replaces the VR `SetterMethodBody`, contained in the `DEFINE` block `SetterMethod` in template `Class` (ll. 19-21 of List. 3). The comments denoting the start and the end of the VR `SetterMethodBody` are defined within a `DEFINE` block, as otherwise the resulting template would be syntactically wrong. This last refinement operation is responsible for introducing assert statements at the beginning of the setter methods. For this purpose, it takes advantage of the `INCLUDE-SUPER` statement, which we introduced to include the original content of the refined `DEFINE` block.

Consequently, the original content is inserted after the assert statement.

```
[DEFINE FurtherMethods FOR MMClass]
  public static [Name] create(...) {
    [FOREACH Attribute AS at]
      assert([at.Name] != null);
    [ENDFOREACH]
    return new [Name](...);
  }
[ENDDEFINE]

[DEFINE Method FOR MMClass]
  [REM]BEGIN VR:SetterMethodBody[ENDREM]
    assert([Name] != null);
    [REM][INCLUDE-SUPER][ENDREM]
  [REM]END VR:SetterMethodBody[ENDREM]
[ENDDEFINE]
```

Xpand

List. 5: Template `ClassWithFact` (in Folder `base`) showing an excerpt of the refining template for the translation of CDs into Java code realized with Xpand.

Please note that, it would not be possible to implement this variability using the XPand language constructs of the used XPand version - only later versions of XPand provide means to customize a code generator. Hence, without our approach, a copy of the original code generator variant would have to be created to develop the shown code generator variant. The decision to use this particular XPand version was rooted in the fact that this version was used in a real-world code generator to which we applied our approach in a case study (see Section 5).

## 5  Industrial Case Study

The approach has been applied to a large real-world Java code generator which processes UML CDs as input. For the contained classes, it generates, among other things, Java classes with mutator and accessor methods. Moreover, each Java class contains additional inner classes and accessor methods that expose the data in a different way and allow a special access to the Java fields. This code generator variant is in the following referred to by *OV1*. Besides this existing code generator *OV1*, a variant of this code generator *OV2* should be build which:

- does not generate the additional inner classes and special access methods.
- does not generate a normal Java field for all UML associations of the corresponding UML class but which generates a field of a special type for UML associations to UML classes tagged with a specific stereotype.
- names the resulting classes according to the originally named classes but with a new suffix, to be able to differentiate between the original and the new classes easily.

The objective of our case study is to demonstrate the usefulness and applicability of our approach to implement a CGPL for a real-world code generator and to compare it to the classical informal approach (copy-paste) for creating CGPLs. For this purpose, we derived the following research questions:

- Is it feasible to apply the approach to establish a CGPL for real-world code generator variants?
- Is the application of the approach superior to the informal reuse of code generators through copy-paste in terms of complexity of the involved artifacts?

## 5.1   Applicability to Real-World Code Generators

In order to better understand the usefulness of the approach, we first implemented the variant *OV2* through informal reuse by doing copy-paste of *OV1*. Then, we applied our approach to realize both generator variants with our approach. For this purpose, we defined a CGL *NV1* which contains the common parts of *OV1* and *OV2*. Moreover, we defined a CGL *NV2* which refines *NV1* in such a way that the generator resulting from the composition of *NV1* and *NV2* generates the same code as *OV1*. Analogously, we defined a CGL *NV3* which refines *NV1* such that the generator resulting from the composition of *NV1* and *NV3* generates the same code as *OV2*. Hence, we assumed that the code generator variants resulting from the composition with the base layer *NV1* must generate the same code as the original code generators - neglecting whitespaces for the sake of simplification.

Using our approach, we were able to derive two code generator variants which generate the same code as code generator variants which did not use our approach. In particular, the presented refinement operations were sufficient to realize the CGPL. For these refinement operations, only `replace` refinement operations have been used, as the developers of the original code generator preferred these over introducing `before` or `after` operations.

## 5.2   Improvements over Informal Reuse

To answer the second research question, we compared the variants *OV1* and *OV2* with the variants *NV1*, *NV2* and *NV3*. To increase comparability, we removed those templates from *OV2* which were copied from *OV1* but not needed for that variant. However, we applied our concept not only on templates, but also on helper classes which can contain more complex functionality which can be accessed from templates. In this use case, helper classes were implemented in Java. The only refinement operation we used in this context was the `replace` operation, which expresses that the implementation of one helper method is replaced by the implementation of another helper method.

To perform the comparison, we measured the templates lines of code (TLOC) and the helper lines of code (HLOC) for *OV1*, *OV2*, *NV1*, *NV2* and *NV3*. To compare our approach with the copy-paste approach, we compared the total TLOC and HLOC of *OV1* and *OV2* with that of *NV1*, *NV2* and *NV3*.

|                      | OV1  | OV2        | $\Sigma_O$    | NV1  | NV2  | NV3 | $\Sigma_N$ |
|----------------------|------|------------|---------------|------|------|-----|------------|
| TLOC                 | 5563 | 2260       | 7823          | 1882 | 4000 | 349 | 6231       |
| Number `DEFINE`      | 327  | 146        | 473           | 189  | 267  | 37  | 493        |
| Number refined `DEFINE` | -  | -          | -             | -    | 94   | 36  | 130        |
| HLOC                 | 929  | 929 (665)  | 1858 (1594)   | 630  | 330  | 49  | 1009       |
| Number helper        | 100  | 100 (79)   | 200 (179)     | 78   | 34   | 5   | 117        |
| Number refined helper | -   | -          | -             | -    | 8    | 8   | 16         |

Tab. 1: Case study results: TLOC and HLOC for the different variants

Table 1 gives an overview over the measured values for the original generator *OV1* and the variant *OV2* created through copy-paste of *OV1* and the generator variants obtained by using our approach. The primary numbers relevant for this comparison are TLOC and HLOC. For *OV2* two HLOC numbers are given: the first results from simple copy-paste of the original helpers, the second number refers to the case that only the helpers used by the variant are counted. Thus, the existing helpers have been analyzed and the helpers not needed were removed. For *NV2* and *NV3*, the number of helpers refers to the number of additionally introduced helper methods. $\Sigma_O$ refers to the sum of the values of both variants *OV1* and *OV2*. Accordingly, $\Sigma_N$ refers to the sum of the values for the variants *NV1*, *NV2* and *NV3*.

As can be seen in Table 1, we can reduce the TLOC size to approximately 79% of the original code generators using our approach. Moreover, we can reduce the HLOC size to approximately 54% respectively 63% of the original code generators.

In addition to that, Table 1 shows that the total number of `DEFINE` blocks is comparable for both variants. Even though `DEFINE` blocks can potentially be reused by multiple variants, this effect does not become apparent in this case, as only two generator variants are created and for each refinement of a `DEFINE` block, one `DEFINE` was introduced, increasing the total number of `DEFINE` blocks. For helper methods, a significant reduction can be observed, as most helper methods can be reused by both generator variants and only few refinements were necessary.

## 6 Related Work

Different annotative, compositional, and transformational modeling approaches have been proposed to express variability in the solution space [Sc12]. Annotative approaches specify all variants in one model. Compositional approaches combine different model fragments to derive a specific variant [HW07, NK08]. Delta modeling [Ha11] applies transformations to a core model. Only few of them have been successfully applied to CGPLs.

In the following, we present existing approaches to address variability in code generators with a special focus on existing code generator frameworks and how they support variability. The concepts we presented are independent of a concrete code generator framework and template language. Another difference to most existing approaches is that arbitrary regions can be marked as VRs.

In the Genesys [JMS08] framework, new generators are established by composing existing Service Independent Building Blocks (SIBs), the atomic unit provided for composition. This approach has been evaluated in many case studies: in most cases, new generators could be derived by the introduction of a small set of new SIBs and a slightly modified composition. This specific mapping represents one point of variation, which can easily be adapted for different targets. The main part of variation are the SIBs, which can be modified via configuration parameters, via a modification of the their execution flow or by replacing a service adapter, which contains execution code for a specific task. In contrast to our approach, Genesys defines a set of different explicit concepts (parameter, service adapters, outgoing branches) to achieve the necessary variation. Our proposed approach of VRs allows to introduce variation points on different kinds of development artifacts and the related `before` and `after` operations can be used to manipulate the execution flow where necessary, too. This way it is also possible to apply variation points to templates, while in [JMS08] templates are modified directly and no variation points are introduced on that level.

[VG07b] highlights the necessity to combine model-to-model transformations and template-based code generation to perform efficient code generation. They suggest that all structural differences on model level should be handled by the transformation layer. [PT02] follow this by pointing out that the generator should handle only two kinds of variation: target variation and the establishment of higher-level primitives based on low-level primitives. Our approach does not provide a guideline on which level which kind of variation should be established, but represents a general concept, to be able to apply variation points where required. If a model-to-model transformation is performed via Java helper classes, corresponding variation can also be applied on that level.

Acceleo [Ac15] provides the concept of dynamic overriding to customize existing generators. To dynamically override templates, a module (which can comprise multiple templates) must extend a module of the existing generator. The extending modules are treated with a higher priority than overridden modules. Thus, the overriding template is invoked instead of the existing template. Templates can only be exchanged as a whole, no variation points can be introduced inside a template.

The template language Xpand supports the customization of code generators using aspect-oriented programming (AOP) [VG07a]. Aspects can be provided which contain template code that is, e.g., invoked instead of code contained in a specific block in the template. Although the original template definition is intercepted, the original overridden template code can still be called in the aspect code [El11]. Our approach is motivated by the concepts applied in Xpand. The main difference to our approach is that our approach does not require support for AOP in the code generator. In Xpand's successor Xtend [Xt15], code generators are composed of extension methods. To customize a code generator written in Xtend, any extension method of a code generator can be exchanged by means of dependency injection. However, these concepts are completely based on language constructs. In contrast, our approach is more general and can be realized with different languages.

## 7 Conclusion

Monolithic code generators are hard to adapt to new requirements and target platforms and, thus, are hardly reusable in different contexts, as they are not designed for adaptations. To overcome this limitation regarding customizations, code generator variability needs to be handled as a primary concern.

We have presented an approach for modeling variability in template-based code generators. This approach relies on *variability regions (VR)* that define extension points in artifacts. Furthermore, since it is an extension of feature-oriented programming, the artifacts are structured in layers that represent code generator features. We additionally define three refinement operations to extend VRs. In order to extend a code generator with a new feature, a new layer can be introduced and existing VRs can be refined. The benefit of the proposed concept is that it is independent of any language that is used for code generator development. We achieve this by introducing a layer definition model language that can be used with any other language. By means of this, the approach facilitates reusing and customizing code generators.